\documentclass[12pt]{article}
\usepackage{amsmath}
\usepackage{latexsym}
\usepackage{amsfonts}
\usepackage[normalem]{ulem}
\usepackage{soul}
\usepackage{array}
\usepackage{amssymb}
\usepackage{mathtools}
\usepackage{extarrows}
\usepackage{graphicx}

\usepackage{subfig}
\usepackage{wrapfig}
\usepackage{wasysym}
\usepackage{enumitem}
\usepackage{adjustbox}
\usepackage{ragged2e}
\usepackage[svgnames,table]{xcolor}
\usepackage{tikz}
\usepackage{longtable}
\usepackage{changepage}
\usepackage{setspace}
\usepackage{hhline}
\usepackage{multicol}
\usepackage{tabto}
\usepackage{float}
\usepackage{multirow}
\usepackage{makecell}
\usepackage{fancyhdr}
\usepackage[toc,page]{appendix}
\usepackage[hidelinks]{hyperref}
\usetikzlibrary{shapes.symbols,shapes.geometric,shadows,arrows.meta}
\tikzset{>={Latex[width=1.5mm,length=2mm]}}
\usepackage{flowchart}\usepackage[paperheight=11.69in,paperwidth=8.26in,left=1.0in,right=1.0in,top=1.5in,bottom=1.5in,headheight=1in]{geometry}
\usepackage[utf8]{inputenc}
\usepackage[T1]{fontenc}
\TabPositions{0.5in,1.0in,1.5in,2.0in,2.5in,3.0in,3.5in,4.0in,4.5in,5.0in,5.5in,6.0in,}
\usepackage[aftersep]{titlesec}
\titlespacing{\subsection}{0pt}{5mm}{1mm}
\renewcommand{\_}{\kern-1.5pt\textunderscore\kern-1.5pt}

\setlistdepth{9}
\renewlist{enumerate}{enumerate}{9}
		\setlist[enumerate,1]{label=\arabic*)}
		\setlist[enumerate,2]{label=\alph*)}
		\setlist[enumerate,3]{label=(\roman*)}
		\setlist[enumerate,4]{label=(\arabic*)}
		\setlist[enumerate,5]{label=(\Alph*)}
		\setlist[enumerate,6]{label=(\Roman*)}
		\setlist[enumerate,7]{label=\arabic*}
		\setlist[enumerate,8]{label=\alph*}
		\setlist[enumerate,9]{label=\roman*}

\renewlist{itemize}{itemize}{9}
		\setlist[itemize]{label=$\cdot$}
		\setlist[itemize,1]{label=\textbullet}
		\setlist[itemize,2]{label=$\circ$}
		\setlist[itemize,3]{label=$\ast$}
		\setlist[itemize,4]{label=$\dagger$}
		\setlist[itemize,5]{label=$\triangleright$}
		\setlist[itemize,6]{label=$\bigstar$}
		\setlist[itemize,7]{label=$\blacklozenge$}
		\setlist[itemize,8]{label=$\prime$}

\begin{document}
\setlength{\parskip}{0pt}
\setlength{\parindent}{1.5em}
\addcontentsline{toc}{section}{Journal Title}
\section*{Semiclassical Effects in Color Flavor Locked \newline Strange Stars}
\addcontentsline{toc}{section}{Semiclassical Effects in Color Flavor Locked Strange Stars}
Guilherme L. Volkmer,\textsuperscript{1} Dimiter Hadjimichef \textsuperscript{1}\par

\textsuperscript{1 }Universidade Federal do Rio Grande do Sul, Porto Alegre/ Av. Bento Gonçalves 9500, Brazil\par

Correspondence should be addressed to Guilherme L. Volkmer; \newline volkmer.guilherme@gmail.com\par

\phantomsection
\subsection*{Abstract}
\addcontentsline{toc}{subsection}{Abstract}
Strange stars in the color flavor locked phase, as described by a generalization of the phenomenological MIT bag model proposed in the context of color superconductivity, are studied through a model motivated by the semiclassical solutions for hydrostatic equilibrium. 
Results show that within this framework it is possible to find  ultracompact configurations situated between regular compact stars and black holes.
\par
\begin{multicols}{2}
\subsection*{1. Introduction}
\addcontentsline{toc}{subsection}{Introduction}
General relativity, the currently accepted theory of the gravitational field, was proposed by Einstein more than a hundred years ago and has been confirmed by all experiments and astrophysical observations so far. Despite the success, general relativity is a completely classical theory, therefore it inevitably imposes a formal discontinuity between gravitational and quantum principles. In fact, Einstein himself was aware that quantum effects would demand modifications in his theory \cite{rovelli}. 

Considering that the subject of quantum gravity remains highly speculative, semiclassical gravity offers a less ambitious  approach from which new insights are feasible. In this framework quantum fields interact with a classical unquantized spacetime metric. Such procedure, as John Earman described, can be seen as a  \textquotedblleft shot-gun marriage\textquotedblright \; between general relativity and quantum field theory, the two pillars of modern physics \cite{handbook}. However, due to the character of its foundations, it is unlikely that such a program reflects an exact description of nature since it combines interactions between quantum fields which are treated in probabilistic terms, with a classical gravitational field relying on well determined values \cite{wald-semi}. Even though some interesting results were derived using semiclassical gravity. The most remarkable example is Hawking's demonstration that black holes can evaporate through a quantum induced thermal radiation,
reducing its mass, and consequently its surface area. 

Studies have shown that semiclassical effects could be relevant even in relativistic stars,
challenging the common view that the influence of gravity on quantum fields is negligible except in the most extreme environments \cite{lima,lima2}. It has also been proposed that the phenomenon known as quantum vacuum polarization in the presence of a gravitational field (where the vacuum state is changed due to the influence of the classical background) may be a new stabilizing ingredient in horizonless ultracompact objects, like black stars, but explicit models for such objects are not currently available \cite{rubio}. In addition, the current gravitational-wave era in astronomy is renewing the interest in theoretical scenarios where new classes of highly compact objects could emerge. For instance, it has been argued that the low mass companion (about 2.6 $M_{\odot}$) of the black hole in the source of GW190814 could be a strange star \cite{ligo,bombaci}.
 
Strange stars are motivated by the so-called strange matter hypothesis, developed independently by Bodmer and Witten, which asserts that the true ground state of the strong interaction is strange quark matter, composed of an approximately equal proportion of up, down, and strange quarks \cite{witten,bodmer}.  Since Witten's work the MIT bag model with unpaired strange quark matter has been widely used to study strange stars. The model considers a gas of free relativistic quarks where confinement is achieved through a vacuum pressure, called the bag constant $B$. 

More sophisticated quark dynamics can be introduced by applying the BCS mechanism to quark matter. 
The strong interaction among quarks is very attractive in some channels and quarks are expected to form Cooper pairs easily (which in quark matter always implies color superconductivity) \cite{weber,alford,rezzola}. Although many pairing schemes have been proposed, if central densities in compact stars are sufficient to support quark matter, it will probably manifest itself through the most symmetrical state, namely the color flavor locked (CFL) phase. 
In this phase all three quarks composing strange quark matter are paired on an approximately equal footing and form a color condensate. When applied to strange stars, It has been shown that the CFL state affects the mass-radius relationship considerably, allowing configurations with large maximum masses \cite{flores}. 

It should be emphasized that there is no clear criteria to establish when semiclassical corrections start being relevant, but some perspective of how extreme a compact star is can be given in terms of the compactness parameter $\mathcal{C}$ (which in $G=c=1$ units is just the ratio between gravitational mass and radius) \cite{rezzola2}.
 For instance, general relativity introduces significant corrections in newtonian gravitation for $\mathcal{C}\approx 0.1$ \cite{shibata}. A CFL strange star described as in general relativity has a compactness parameter which is more than half of a black hole for all nineteen parametrizations proposed in Ref. \cite{flores}. Although CFL strange stars are not as extreme as some hypothetical models available in the literature, they still offer a high density environment combined with a solid theoretical background, avoiding the problems usually found in more extreme proposals \cite{rubio}.

All things considered, it seems pertinent to evaluate the impact of semiclassical effects in CFL strange stars and the rest of the paper is organized as follows: In Section 2 basic aspects regarding semiclassical hydrostatic equilibrium are discussed and some appropriate limits for dealing with strange stars are established. It will be hypothesized that the environment produced by  CFL strange stars is able to ignite a semiclassical correction, absent in general relativity,  that will be associated with a generic linear equation of state with negative pressure. This is inspired in Gliner's idea that certain gravitational collapses might reach an ultradense \textquotedblleft vacuum-like\textquotedblright\; state, that could be described phenomenologically by an negative pressure equation of state \cite{gliner}. Section 3 is devoted to the equation of state of CFL quark matter and the set of parametrizations that will be adopted further. Section 4 presents the mass-radius relations for the proposed model and compares its results with those obtained via general relativity. Section 5 closes the article with the final remarks.  
\subsection*{2. Semiclassical Hydrostatic Equilibrium}
\addcontentsline{toc}{subsection}{Semiclassical Hydrostatic Equilibrium}
Aiming to go beyond the generally relativistic picture regarding hydrostatic equilibrium, semiclassical gravity offers a conservative framework which should be valid in many situations where the spacetime curvature is not comparable to the Planck length.
So in what follows only situations where the fluctuations of the gravitational field are negligible will be considered. In this case a classical metric $g_{\mu\nu}=\langle \hat{g}_{\mu\nu}\rangle$ is assumed just as in general relativity \cite{frolov}.\par
Now suppose that this classical spacetime is populated by a collection of quantum fields assumed to be in a given quantum state. In this context a semiclassical version of the Einstein field equations is proposed replacing the classical energy-momentum tensor by the expectation value of the energy-momentum tensor of the relevant quantized fields in the chosen state \cite{visser}.
Particularly, using the $1/N$ expansion, one assumes that there are $N$ non-interacting scalar fields present and the 
 semiclassical Einstein equations can be expressed as \cite{rubio}
\begin{equation}
 G_{\mu\nu}=8\pi \left( G T_{\mu\nu}+\hbar N \Theta Q_{\mu\nu}\right)+\mathcal{H}_{\mu\nu} \label{sceee},
\end{equation}
where $Q_{\mu\nu}$ represents the main contribution due to quantum effects and $\Theta$ is a constant.
The term $\mathcal{H}_{\mu\nu}$ represents subdominant contributions and can be safely neglected for the present purposes \cite{rubio}.

The subsequent analysis is restricted to static and spherically symmetric spacetimes, with line element of the following form
(hereafter Greek indices take four values and latin indices only two)
\begin{align}
 ds^{2}&=ds^{2}_{(2)}+r^{2}d\Omega^{2}\nonumber\\
 &=g_{ab}(y)dy^{a}dy^{b}+r^{2}(y)d\Omega^{2}(\theta, \phi),\label{le}
\end{align}
where $d\Omega^{2}(\theta, \phi)$ denotes the angular line element on the 2-sphere.

The classical source in (\ref{sceee}) will be described by a perfect fluid mathematically represented by
\begin{equation}
 T_{\mu\nu}=\left(\varepsilon+p\right)u_\mu u_\nu +pg_{\mu\nu}\label{cset},
\end{equation}
where $p$ is the pressure, $\varepsilon$ is the energy density and $u^{\mu}$ (satisfying $u_\mu u^\mu=-1$) denotes the components of the four-velocity.\par
Regarding the semiclassical correction, it is useful to remember that in curved spacetime there
is no preferred “vacuum state”. For instance,
if the gravitational field contains an event horizon, there are at least four different natural definitions of the quantum mechanical vacuum state \cite{visser-gvp}. Fortunately, for a horizonless compact star of mass $M$ in empty space, a more clear scenario rises and the appropriate vacuum state is known to be the Boulware vacuum state, a state with zero particle content for static observers and regular both inside and outside the star, also known as Schwazschild vacuum. This vacuum state is approximately the same as the Minkowski vacuum at $r\gg 2M$. 

Furthermore, since there are no exact analytical expressions available for $Q_{\mu\nu}$ in four dimensions, the s-wave Polyakov approximation is also assumed \cite{rubio}. The basic idea is that, under spherical symmetry, a decomposition through spherical harmonics allows the reduction from a four-dimensional theory to a set of two-dimensional theories specified by the angular-momentum. The dominant contributions in some spherically symmetric systems come from the “s-wave sector”, that is, the $l=0$ mode \cite{balbinot}. 
In physical terms it is responsible for neglecting quantum fluctuations that are not spherically symmetric and also effects of back-scattering \cite{rubio}. In this approximation, the dimensional reduction is given by 
 \begin{equation}
   Q_{\mu\nu} = \frac{\delta^{a}_{\mu}\delta^{b}_{\nu}}{4\pi r^{2}} Q^{\left(2\right)}_{ab}\label{qset},
\end{equation}
which is calculated with respect to $ds^{2}_{(2)}$ as expressed in equation ($\ref{le}$). Fortunately, $ Q^{\left(2\right)}_{ab}$ can be expressed in a completely tensorial way as \cite{rubio}
 \begin{equation}
    Q^{\left(2\right)}_{ab}=\frac{1}{48\pi}\left(R^{(2)}g_{ab}+A_{ab}-\frac{1}{2}g_{ab}A\right)\label{tensorial},
 \end{equation}
 where $A_{ab}= 4|\partial_t|^{-1} \nabla_a\nabla_b |\partial_t|$ for the Boulware state. \par
 Therefore hydrostatic equilibrium in semiclassical gravity is determined by solving the semiclassical Einstein equations as expressed in (\ref{sceee}) for the sources described by equations (\ref{cset}) and (\ref{qset}) under spherical symmetry, yielding \cite{rubio}
\begin{equation}
 p^{\prime}\left(1-\frac{ p^{\prime}}{2\Upsilon(r)}\right)=\Xi(r)\label{she},
\end{equation}
where for any function a prime denotes a derivative with respect to r. For simplicity it was defined:
\begin{equation}
 \Xi(r)\coloneqq -\frac{G}{r}\frac{\left(\varepsilon+p\right)\left(m+4\pi r^{3}p\right)}{r-2Gm},
\end{equation}
which represents the solution obtained in general relativity, and
\begin{equation}
    \Upsilon(r)\coloneqq\frac{12\pi r\left(p+\varepsilon\right)}{\hbar \Theta N},
\end{equation}
carrying the semiclassical influence in hydrostatic equilibrium.
Equation (\ref{she}), being quadratic with respect to the pressure gradient, has two roots given by
\begin{equation}
     p^{\prime}_\pm =\Upsilon(r)\left(1\pm\sqrt{1-\frac{2\Xi(r)}{\Upsilon(r)}}\right).\label{pmr}
\end{equation}
Exact solutions for $p^{\prime}_{+}$ were analyzed in Ref. \cite{rubio} by assuming that the quantity inside the square root is equal to a constant. The picture which emerged appears to be a nontrivial combination of two hypothetical compact objects, namely, black stars and gravastars \cite{rubio}. Here a different approach will be adopted. 
First, observe that when $\hbar \to 0$, $p^{\prime}_{-}$ reduces to
\begin{equation}
    p^{\prime}_{-} =\Xi(r) \label{ctov},
\end{equation}
that is, at the classical limit the usual form of the TOV equation is recovered \cite{rubio}. 

Additionally, it can be shown that the semiclassical mass function adds a term proportional to $\hbar$ to the expression known from general relativity \cite{rubio}. So it also recovers the classical expression when $\hbar \to 0$, namely
\begin{equation}
    m^{\prime}=4\pi r^{2}\varepsilon \label{grmf}.
\end{equation}
\par
On the other hand, the $p^{\prime}_{+}$ solution does not present a classical limit. Nevertheless it would be interesting for astrophysical applications to obtain some expression that operationally resembles a classical limit. Observe that $p^{\prime}_{+}$ can be rewritten as:
\begin{align}
    \Theta\hbar p^{\prime}_{+}=&\frac{12\pi r\left(\varepsilon+p\right)}{N}\nonumber\\
    &\left(1+\sqrt{1-\frac{\Theta\hbar N}{6\pi r \left(\varepsilon+p\right)}\Xi}\right).
\end{align}
Through a Taylor series about $\hbar=0$ and matching both sides in powers of $\hbar$ the above equation yields
\begin{equation}
   p^{\prime}_{+}=-\Xi(r) =\frac{G}{r}\frac{\left(\varepsilon+p\right)\left(m+4\pi r^{3}p\right)}{r-2Gm}\label{lnl}.
\end{equation}
Since the number $N$ is essentially arbitrary,
the other terms in the expansion do not impose any restrictions, vanishing identically, once the limit $N \to \infty$ with $\Theta N$ fixed is taken in all of them \cite{hartle}. This strategy if applied to the other root would give again as result the classical TOV equation.

Interesting to notice that equation (\ref{lnl}) differs from the TOV equation only by a minus sign. It also has to fulfill the condition $r>2Gm$ at any point inside the star, thus forbidding the presence of a Schwarzschild black hole at any radius r. Now observe that, for ordinary forms of matter with $\varepsilon >0$ and $p>0$, equation (\ref{ctov}) describes a negative pressure gradient 
and the pressure is a monotonically decreasing function of the radius which eventually vanishes. It is not difficult to realize that in order to have a similar picture regarding equation (\ref{lnl}), unconventional forms of matter should be employed. Here, equation (\ref{ctov}) will be associated with an equation of state with negative pressure of the form 
\begin{equation}
  p=-\omega\varepsilon,  \label{npeos}
\end{equation}
where $\omega$ is a positive parameter smaller than unity. For configurations where $m>|4\pi r^{3}p|$ the pressure gradient is positive and assures that the absolute value of the pressure (and also the energy density) decreases as the radius increases. Physically speaking, equation (\ref{npeos}) is responsible for phenomenologically taking into account quantum gravitational polarization effects, modifying the predictions for quantities such as mass and radius of compact objects.
\subsection*{3. The Color Flavor Locked Equation of State}
\addcontentsline{toc}{subsection}{Color Flavor Locked Strange Stars}
The CFL equation of state is a nonlinear generalization of the unpaired version of the MIT bag model, proposed in the context of color superconductivity. 
Different parametrizations are possible depending on the values of $B$ (the bag constant), $\Delta$ (the gap of the QCD
Cooper pairs) and $m_S$ (the strange quark mass), which are not accurately known and are taken as free parameters. All results in the next section refer to the set of parametrizations presented in Ref. \cite{flores}, which are also displayed in Table \ref{tab1}. 

It is customary to use a semi-empirical model in which the thermodynamic potential to order $\Delta^{2}$ can be expressed as \cite{paulucci}
\begin{equation}
    \Omega_{CFL}=\Omega_{free}-\frac{3}{\pi^{2}}\Delta^{2}\mu^{2}+B,
\end{equation}
where $\Omega_{free}$ represents the non-paired state and $\mu$ is the chemical potential.
From the above equation, pressure and energy density can be analytically expressed to order $m_s^{2}$ as
\begin{equation}
    p=\frac{3\mu^{4}}{4\pi^{2}}+\frac{9\alpha\mu^{2}}{2\pi^{2}}-B,
    \end{equation}
    \begin{equation}
    \varepsilon= \frac{9\mu^{4}}{4\pi^{2}}+\frac{9\alpha\mu^{2}}{2\pi^{2}}+B,
\end{equation}
 with
\begin{equation}
    \alpha=-\frac{m_{S}^{2}}{6}+\frac{2\Delta^{2}}{3}.
\end{equation}
It is straightforward to express pressure and energy density as a one parameter equation of state of the form $\varepsilon(p)$, namely
\begin{equation}
    \varepsilon=3p+4B-\frac{9\alpha\mu^{2}}{\pi^{2}},
    \end{equation}
    with the chemical potential expressed by
    \begin{equation}
    \mu^{2}=-3\alpha+\left[\frac{4\pi^{2}}{3}\left(B+p\right)+9\alpha^{2}\right]^{\frac{1}{2}}.
\end{equation}
Alternatively, in a similar fashion $p(\varepsilon)$ is given by
\begin{equation}
    p=\frac{\varepsilon}{3}-\frac{4B}{3}+\frac{3\alpha\mu^{2}}{\pi^{2}},
    \end{equation}
where the chemical potential is expressed in terms of the energy density, namely
    \begin{equation}
    \mu^{2}=-\alpha+\left[\frac{4\pi^{2}}{9}\left(\varepsilon-B\right)+\alpha^{2}\right]^{\frac{1}{2}}.
\end{equation}
\begin{table}[H]
\Huge
\resizebox{\columnwidth}{!}{%
\begin{tabular}{|c|c||c|c|}
\hline
Parametrization & $B\left(MeV/fm^3\right)$ & $\Delta(MeV)$ & $m_S(MeV))$\\
\hline
CFL1 & 60 & 50 & 0\\
CFL2 & 60 & 50 & 150\\
CFL3 & 60 & 100 & 0\\
CFL4 & 60 & 100 & 150\\
CFL5 & 60 & 150 & 0\\
CFL6 & 60 & 150 & 150\\
CFL7 & 80 & 100 & 0\\
CFL8 & 80 & 100 & 150\\
CFL9 & 80 & 150 & 0\\
CFL10 & 80 & 150 & 150\\
CFL11 & 100 & 50 & 0\\
CFL12 & 100 & 100 & 0\\
CFL13 & 100 & 100 & 150\\
CFL14 & 100 & 150 & 0\\
CFL15 & 100 & 150 & 150\\
CFL16 & 120 & 100 & 0\\
CFL17 & 120 & 150 & 0\\
CFL18 & 120 & 150 & 150\\
CFL19 & 140 & 150 & 0\\
\hline
\end{tabular}
}
\caption{\label{tab1} Parametrizations originally presented in Ref. \cite{flores} which are also adopted in this work.}
\end{table}
Among its features is the fact that the CFL phase is a candidate for the true ground state of hadronic matter for a  much wider range of the parameters of the model than the state without any pairing \cite{flores,lugones}. Another property is that as the gap increases the equation of state gets stiffer, allowing configurations with higher maximum masses \cite{paulucci}.
 In the next section the impact of the semiclassical correction in the mass-radius relationship of CFL strange stars will be analyzed. 

\subsection*{4. Results and Discussion}
The system of hydrostatic equilibrium equations, as expressed in (\ref{ctov}) and (\ref{lnl}), will be applied to a strange star where both contributions interact only gravitationally. The mass function for both cases is given by the equation (\ref{grmf}) as in general relativity. Numerical solutions are obtained similarly to standard procedure in general relativity, keeping in mind that here each fluid satisfies its own hydrostatic equilibrium equation. The system is solved from the center, with $m(r=0)=0$ for both contributions, and the energy densities are needed as input.  Here, in order to restrict the role of semiclassical effects in the model, the central energy density associated with the equation (\ref{npeos}) is taken to be $10\%$ of the main contribution coming from the CFL phase. The integration stops when the effective pressure vanishes, defining the final mass (which is the total gravitational mass from its two contributions) and radius. It is worth mentioning that although the CFL strange star will be embedded in a negative pressure environment, all solutions considered have an non-negative effective pressure and energy density.

\begin{figure}[H]
\begin{center}
\includegraphics[scale=0.45]{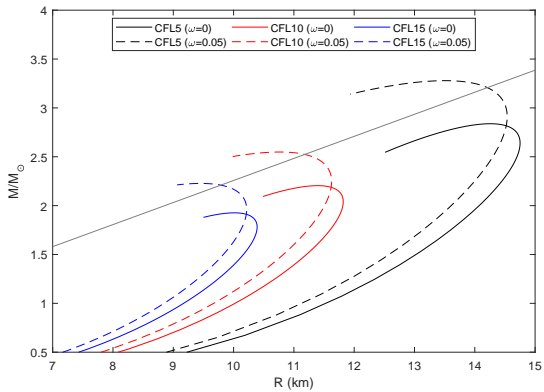}
\caption{Comparison of the mass radius relationship obtained in general relativity and semiclassical gravity for some parametrizations of the CFL equation of state.}
\label{fig1}
\end{center}
\end{figure}

\begin{figure}[H]
\begin{center}
\includegraphics[scale=0.45]{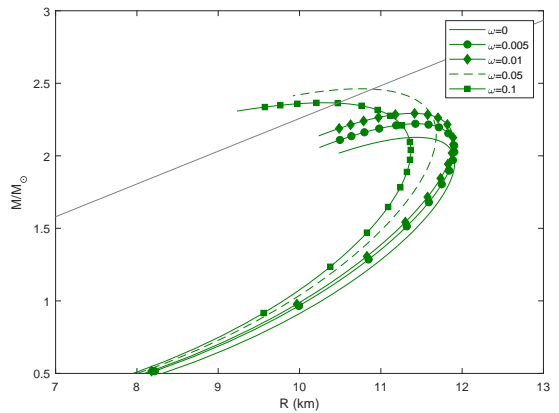}
\caption{Impact of the $\omega$ parameter on the mass-radius curve for the parametrization CFL4.}
\label{fig2}
\end{center}
\end{figure}

As illustrated in Figures \ref{fig1} and \ref{fig2}, the main feature of introducing the semiclassical correction is the possibility of finding ultracompact configurations (defined by $\mathcal{C}>1/3$) throughout all parametrizations without imposing significant deviations in the low mass-radius region. 
In Figures \ref{fig1} and \ref{fig2}, a grey line corresponding to $\mathcal{C}=1/3$ was added to demarcate the ultracompact region.
 Specifically, Figure \ref{fig1} confronts the mass-radius curve obtained using general relativity and semiclassical gravity for some parametrizations of the CFL equation of state, while the $\omega$ parameter is fixed in $0.05$ (in $G=c=1$ units where the parameter is dimensionless). The semiclassical configurations are more massive and smaller, crossing to the ultracompact region. 
On the other hand, in Figure \ref{fig2} the $\omega$ parameter is varied while the parametrization is fixed. The goal is to illustrate that the maximum mass can not be increased indefinitely within the model. Due to the negative pressure associated with the component carrying the semiclassical corrections, after some value of $\omega$ that depends on the inputs for the central energy densities, the solutions become less massive. 
Table \ref{tab2} summarizes the results for the nineteen parametrizations of the CFL equation. The data corresponds to the  maximum mass configurations. In order to interpret the implications of such solutions is interesting to observe that semiclassical gravity and general relativity can produce, for different parametrizations, similar results for mass and radius (compare for example the CFL9 parametrization in general relativity and the CFL1 parametrization in semiclassical gravity).
Probably a mass measurement would not be able to distinguish among such objects, even though they would be physically different. 
The ultracompact regime exhibits a variety of interesting visual effects when compared to regular
compact stars \cite{nemiroff,nemiroff2}. A critical distinctive feature is the presence of a photon sphere, in other words, the unstable
circular null geodesic of the external Schwarzschild spacetime metric \cite{urbano}. Ultracompact stars may also play
an important role if future gravitational wave data confirm the phenomenon known as gravitational echoes, which are secondary pulses of gravitational radiation after the main burst of radiation related to the post-merger ringdown waveform. 
It has been argued that gravitational echoes are not, as commonly assumed, a unique prerogative of deviations from general relativity at the horizon
scale \cite{cardoso,pani}. Similar signals may arise from  ultracompact stars and could be a powerful physical property to distinguish different compact objects,
possibly revealing a new branch of stable configurations. First, echoes imply that the remnant is more compact than a neutron star with an ordinary equation of state \cite{pani}. In addition, although the post-merger ringdown waveform of an ultracompact star is initially identical to that of a black hole, the echoes in the late-time ringdown would present significant differences \cite{pani2}. Therefore at least in principle ultracompact stars could be distinguished from both ordinary neutron stars and black holes. 
\begin{table}[H]
\Huge
\resizebox{\columnwidth}{!}{%
\begin{tabular}{|c|c||c|}
\hline
 Parametrization & General Relativity & Semiclassical Gravity\\
  & ($M(M_{\odot})$, $R(km)$, $\mathcal{C}$) & ($M(M_{\odot})$, $R(km)$, $\mathcal{C}$) \\
\hline
CFL1 & (2.051, 11.08, 0.27) & (2.375, 10.43, 0.34)\\
CFL2 & (1.830, 10.09, 0.27) & (2.120, 9.47, 0.33)\\
CFL3 & (2.357, 12.38, 0.28) & (2.725, 11.68, 0.36)\\
CFL4 & (2.127, 11.41, 0.27) & (2.462, 10.75, 0.34)\\
CFL5 & (2.842, 14.24, 0.29) & (3.278, 13.51, 0.36)\\
CFL6 & (2.631, 13.46, 0.29) & (3.039, 12.74, 0.35)\\
CFL7 & (1.994, 10.52, 0.28) & (2.309, 9.95, 0.34)\\
CFL8 & (1.821, 9.79, 0.27) & (2.111, 9.22, 0.34)\\
CFL9 & (2.365, 11.98, 0.29) & (2.735, 11.36, 0.35)\\
CFL10 & (2.202, 11.36, 0.29) & (2.548, 10.76, 0.35)\\
CFL11 & (1.571, 8.51, 0.27) & (1.823, 8.01, 0.34)\\
CFL12  & (1.754, 9.29, 0.28) & (2.034, 8.79, 0.34)\\
CFL13 & (1.616, 8.70, 0.27) & (1.875, 8.20, 0.34)\\
CFL14 & (2.055, 10.49, 0.29) & (2.379, 9.95, 0.35)\\
CFL15 & (1.922, 9.98, 0.28) & (2.227, 9.44, 0.35)\\
CFL16 & (1.582, 8.40, 0.28) & (1.835, 7.95, 0.34)\\
CFL17 & (1.834, 9.42, 0.29) & (2.125, 8.94, 0.35)\\
CFL18  & (1.722, 8.98, 0.28) & (1.997, 8.51, 0.35)\\
CFL19  & (1.667, 8.60, 0.29) & (1.934, 8.16, 0.35)\\

\hline
\end{tabular}
}
\caption{\label{tab2} Comparison between the maximum mass configurations using general relativity and semiclassical gravity. For the semiclassical solutions it was adopted $\omega=0.05$.}
\end{table}
 
 In Ref. \cite{bombaci} it was discussed the possible tension between the evidence of the existence compact stars satisfying $R \lesssim 11.6\;km$ at $1.4 M_{\odot}$ (suggested by some analyses on thermonuclear bursts and X-ray binaries), and the possibility of very massive stars with $M\sim 2.6 M_{\odot}$. None of the parametrizations can accommodate, using general relativity, a family satisfying $R_{1.4} \lesssim 11.6km$ and also stars with masses as high as $2.6 M_{\odot}$. For the semiclassical solutions this is not necessarily the case. The CFL3 parametrizations, for instance, predicts a radius of $11.40$ km for a star with $M=1.4 M_{\odot}$ and a maximum of $2.725 M_{\odot}$, being able to deal with both conditions at once. 

\subsection*{5. Conclusions}
\addcontentsline{toc}{subsection}{Conclusions}
The present work has explored the implications of incorporating semiclassical effects in a specific compact star model based on known physics, namely, CFL strange stars.
The generalization studied here is in some sense in tune with Einstein's doubts about the reality of his field equations in the face of quantum physics, particularly the right hand side (a phenomenological representation of matter), while he believed that the left hand side (obtained from first principles using geometrical quantities) contained a deeper truth.
To express this contrast he even used to say that the first was made of low grade wood and the second of fine marble.

 Results have shown the possibility of horizonless ultracompact configurations, a class that has received much attention in the literature \cite{rubio,urbano,pani,pani2}. These results were achieved without imposing drastic modifications on the low mass-radius regime. Within the semiclassical framework it is also possible to find families of stars satisfying $R_{1.4} \lesssim 11.6\;km$ and masses as high as $2.6 M_{\odot}$, which is very difficult within regular neutron star or even strange star models \cite{bombaci}.

Conceptually speaking, since any object that undergoes complete gravitational collapse passes through an ultracompact phase, any new effect not taken into account in general relativity could have a prominent role, avoiding the full collapse in certain scenarios. Therefore the results presented here are also pleasing in the sense that the semiclassical analysis enabled an intermediate class of ultracompact stars between regular compact stars and black holes.\par This also establishes an interesting link regarding physical properties, since a neutron star has neither a photon sphere or a event horizon and a black hole has both. Ultracompact stars could be an intermediate step presenting a photon sphere but no event horizon \cite{nemiroff,nemiroff2}.

Events like the LIGO announcement of the discovery of an object inside the so-called mass gap, or even the tentative detection of echoes in the post-merger signal of
the neutron-star binary coalescence GW170817 (with 4.2$\sigma$
significance level), are very stimulating, and new classes of compact stars may be soon a reality \cite{ligo,ligo2}.

\subsection*{Acknowledgments}
\addcontentsline{toc}{subsection}{Acknowledgments}
This work was partially supported by Conselho Nacional de Desenvolvimento Científico e Tecnológico – CNPq.\par

\end{multicols}
\end{document}